\documentclass[apj]{emulateapj}

\newcommand{\ha}{H\ensuremath{\alpha}}
\newcommand{\hi}{H{\scshape i}}
\newcommand{\hii}{H{\scshape ii}}
\newcommand{\nii}{[N{\scshape II}]}   
\newcommand{\oii}{[O{\scshape II}]}   

\submitted{Accepted by ApJ Letters 24 December 2009}

\shortauthors{Guti\'{e}rrez \& Beckman}
\shorttitle{Distributions of mean electron density in HII regions.}

\begin{document}

\title{The galaxy-wide distributions of mean electron \\
       density in the HII regions of M51 and NGC 4449.}

\author{Leonel Guti\'{e}rrez\altaffilmark{1,2} and John E. Beckman\altaffilmark{2,3} }
\altaffiltext{1}{Universidad Nacional Aut\'onoma de M\'exico, Instituto de Astronom\'{\i}a, Ensenada, B. C. M\'exico} 
\altaffiltext{2}{Instituto de Astrof\'{\i}sica de Canarias, C/ Via L\'{a}ctea s/n, 38200 La Laguna, Tenerife, Spain} 
\altaffiltext{3}{Consejo Superior de Investigaciones Cient\'{\i}ficas, Spain}

\email{leonel@astrosen.unam.mx, jeb@iac.es}

\begin{abstract}
Using ACS-HST images to yield continuum subtracted photometric maps in \ha{} of the Sbc galaxy M51 and the dwarf irregular galaxy NGC 4449, we produced extensive (over 2000 regions for M51, over 200 regions for NGC4449) catalogues of parameters of their \hii{} regions: their \ha{} luminosities, equivalent radii and coordinates with respect to the galaxy centers. From these data we derived, for each region, its mean luminosity weighted electron density, $\left<n_e\right>$, determined from the \ha{} luminosity and the radius, $R$, of the region. Plotting these densities against the radii of the regions we find excellent fits for $\left<n_e\right>$ varying as $R^{-1/2}$. This relatively simple relation has not, as far as we know, been predicted from models of \hii{} region structure, and should be useful in constraining future models.  Plotting the densities against the galactocentric radii, $r$, of the regions we find good exponential fits, with scale lengths of close to 10 kpc for both galaxies. These values are comparable to the scale lengths of the \hi{} column densities for both galaxies, although their optical structures, related to their stellar components are very different. This result indicates that to a first approximation the \hii{} regions can be considered in pressure equilibrium with their surroundings. We also plot the electron density of the \hii{} regions across the spiral arms of M51, showing an envelope which peaks along the ridge lines of the arms.
\end{abstract}


\section{Introduction}\label{sec:intro}

Electron densities for \hii{} regions can be measured in two different ways, which respectively 
yield two different ranges of values. Density sensitive emission line ratios, such 
as the [SII] $\lambda\lambda$6731/6716\AA~doublet ratio, or the \oii{}  $\lambda\lambda$3729/3726\AA~doublet ratio give 
values in the range of hundreds \citep{zaritsky94,mccall85}. However measurements using \ha{} surface brightness to 
derive the emission measure, together with the estimated radius of an \hii{} region, give 
values in the range 1-10 \citep[see e.g.][]{rozas96}. This type of 
dichotomy was recognized long ago by \citet{osterbrock59}  who 
used the emission line ratio of the \oii{} doublet, $\lambda\lambda$3729/3726\AA, as their version of the 
first technique, and cm wavelength radio emission (from the literature) to estimate the 
emission measure, i.e. for the second technique. They reached the conclusion that the best
 way to reconcile the differences, of more than an order of magnitude between the two 
techniques, is to assume that the emission nebula is clumpy, with dense clouds embedded 
in a much more tenuous substrate. The line ratios are strongly weighted by density, and 
yield values reflecting the densities of the dense clumps, while the electron density 
derived using the emission measure and the overall size of the \hii{} region is a geometrical 
average, whose balance between the contribution of the clumps and the substrate depends 
on the fractional volume occupied by the former. In models for \hii{} regions based on 
recognition of this inhomogeneity, the fractional volume occupied by the clumps is 
termed the filling factor, and a first order approximation can be made that the 
interclump substrate makes a negligible contribution to the line emission, so that 
the interclump volume can be considered to have negligible electron density (see 
\citet{osterbrock89} for a 
standard treatment). In the present article we are using the measured luminosities of
\hii{} regions in \ha{} to compute an electron density using the second method, so our 
values are to be taken as the means produced by the second technique. We will use 
them to derive comparative electron densities of selected \hii{} regions in M51 and 
NGC 4449, to establish functional relations between them, the sizes of the \hii{} regions,
and their positions within the discs of their respective galaxies, and hence to 
provide an overview of the global behavior of the electron density within the 
discs of the two galaxies.

\section{The observational data and its treatment}\label{sec:data}
M51 was observed in January 2005 with the ACS on HST using the broad band filters 
F435W (close to the standard B band) F555W (close to V) and F814W (close to I) 
and the narrow band filter F658N (\ha{}) within observational program 10452 
(P.I. S. Beckwith, Hubble Heritage Team). The data used here were corrected, 
calibrated, and combined into a mosaic of the galaxy by \citet{mutchler05}.
The images of NGC 4449 were also taken 
with the ACS, through the same filters as those used for M51, in November 2005, 
within program GO 10585 (P.I. A. Aloisi). The images were taken at two different 
positions along the major axis of the galaxy, with four dithered pointings on
 each position \citep{annibali08} to help cosmic ray removal. We combined 
them using MULTIDRIZZLE \citep{koek02b}, yielding overall exposure 
times of 3600s, 2400s, 2000s and 360s respectively for the four named filters. 
The total resulting field is sized 345$\times$200 arcsec$^2$; we produced mosaics 
containing complete images of the galaxy in all bands. 

For both galaxies 
standard correction and calibration procedures were used (bias, flat-field, 
dark current, and distortion corrections) using the ACS pipeline calibration 
suite CALACS \citep{hack00}. Use of the Multidrizzle process 
allowed us to correct for detector defects and distortions, though we 
retained the original pixel scales (0.05 arcsec per pixel). After this 
process we converted the images from e-/s to flux in erg cm$^{-2}$ s$^{-1}$ \AA$^{-1}$ 
using the standard PHOTOFLAM conversion factor \citep{sirianni05}.
We aligned the continuum and \ha{} images and 
produced a continuum subtracted \ha{} image, after deriving the factor of 
proportionality between the continuum in the on-band and off-band filters, 
following the method of \citet{boker99}. A more detailed description of our image 
treatment can be found in Guti\'errez et al. (2010, in preparation). The 
method for measuring 
the \ha{} flux of an \hii{} region requires defining its boundaries in the presence 
of any local diffuse \ha{} emission and any overlapping neighbour regions, 
using circular or rectangular defining apertures. The boundary of an HII region 
is defined as containing only pixels whose brightness is greater
than 3$\sigma$, measured on the background. For NGC 4449 the diffuse 
background subtraction presented special difficulties due to its strength, 
and to perform this accurately we used a selective filter, as defined in 
Guti\'errez et al. (2010, in preparation). Once a local background brightness 
was determined, it was multiplied by the number of pixels within the \hii{} region 
and subtracted  from the flux directly measured within the region, to yield the 
measured flux from the region.  To allow for the overlap of regions,  if the 
brightness between the two maxima falls below 2/3 of the value of the fainter 
maximum we considered them to be two separate regions, with the boundary in the 
depression of the brightness distribution. Where this criterion was not met we 
classified the region as single.

To allow for 
the contribution of the \nii{} doublet through the F658N filter, for M51 
we used the mean of the observed ratios of these lines to \ha{} for 10 \hii{} regions in 
that galaxy by \citet{bresolin04}, which show very limited variations
from region to region, to calculate the factor by which the observed 
flux should be reduced, using the filter transmission curve and the
observed redshift of M51. For NGC 4449 we used the observed ratio from 
\citet{kennicutt92}.
The factor for M51 was 0.67, and the corresponding factor for NGC 4449 is 0.86.
We understand that radial decline in metallicity
(especially in M51), will give rise to errors in using these blanket numbers
 for each object, but we can show that these errors are of second order 
(Guti\'errez et al. 2010, in preparation). Also, we estimated the possible systematic error due to 
the presence of the [OIII] emisison line 
at $\lambda$5007\AA~in the F555W filter and found this to be less than 1\%. 

Finally, in deriving absolute fluxes we took 
the distance to M51 as 8.39 Mpc, following \citet{feldmeier97} and the 
distance to NGC 4449 as 3.82 Mpc, following \citet{annibali08}. We 
measured the luminosity of each region as outlined above, and also, from 
the continuum- and background-subtracted \ha{} image, determined the area 
subtended by the image, and derived a value for an equivalent radius, 
obtained dividing the area by $\pi$ and taking the square root. With these 
measurements we could determine a mean electron density for a each 
region, as described below. 
A catalogue of 2657 regions of M51, and 273 regions of NGC4449 was produced, 
containing positions, measured fluxes and luminosities in \ha{}, and equivalent 
radii (see Guti\'errez et al. 2010 for details). 



\section{The mean electron density of the H\small{II} regions}\label{sec:density}
\subsection{Electron density as a function of H\small{II} region radius}

We described in the introduction the inhomogeneity of \hii{} regions, but 
with the set of observations described here we confine our attention to 
the mean electron density ($\left< n_e\right>$), and use the ``Case B'' formula \citep{osterbrock89}

\begin{equation}
\left<n_e\right> = \sqrt{\frac{2{.}2 L({\ha})}{\frac{4}{3}\pi\alpha_B{h \nu_{\ha{}}}R^3}} \label{eq:n1}
\end{equation}

\noindent to derive this, where L(\ha{}) is the \ha{} luminosity, $\alpha_B$ is the case B 
recombination coefficient, $h \nu_{\ha{}}$ is the energy of an \ha{} photon, and $R$ 
is the equivalent radius of the \hii{} region. If we measure $R$ in pc and L(\ha{}) 
in erg s$^{-1}$, and taking a canonical value for the temperature of 10000 K we can 
rewrite equation \ref{eq:n1} as

\begin{equation}
\left<n_e\right> = 1{.}5 \times 10^{-16} \sqrt{\frac{L({\ha{}})} {R^3}} cm^{-3}.\label{eq:n2}
\end{equation}

\begin{figure}
\centering
   \includegraphics[width=3.1in]{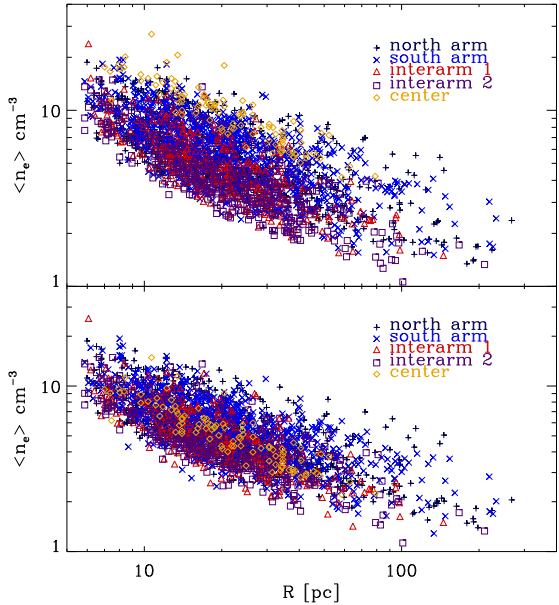} 
   \caption{Mean electron densities of all the \hii{} regions 
measured in M51 as a function of equivalent radius. In the lower panel the densities have been corrected by a factor which takes into account their galactocentric distances (plotted in Fig. \ref{F:densidad_vs_distancia_m51}).} \label{F:densidad_vs_radio_m51} \end{figure}

\begin{figure}
\centering
   \includegraphics[width=3.1in]{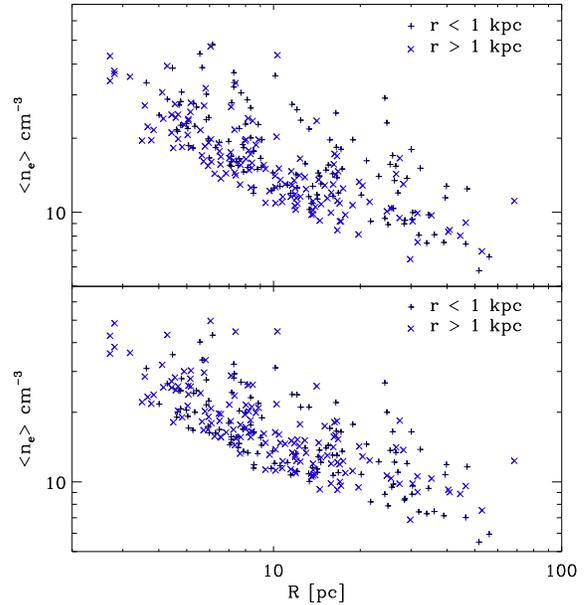} 
   \caption{Mean electron densities of the \hi{} regions 
measured in NGC 4449 as a function of equivalent radius. In the lower panel the densities have been corrected by a factor which takes into account their galactocentric distances (plotted in Fig. \ref{F:densidad_vs_distancia_ngc4449}).} \label{F:densidad_vs_radio_n4449} 
\end{figure}

In Fig.~\ref{F:densidad_vs_radio_m51} we show the mean electron densities of all the regions 
observed in M51 against their equivalent radii. There is considerable scatter, 
which we will investigate further below. A linear fit to the points gives the 
result:

\begin{equation}
\left<n_e\right>  = \frac{28{.}9}{R^{0{.}56}} \label{eq:n3};
\end{equation}

\noindent in qualitative terms, the larger regions have lower mean electron densities. 
The functional trend is quite well reproduced by the \hii{} regions in NGC 4449 
as we can see in Fig. \ref{F:densidad_vs_radio_n4449}. With far fewer points, the best linear 
fit is given by 

\begin{equation}
\left<n_e\right>  = \frac{44{.}2}{R^{0{.}45}} \label{eq:n4}.
\end{equation}

These are only two galaxies, but the trends can at this stage be usefully 
summarized by the simple expression

\begin{equation}
\left<n_e\right>  \sim  \frac{1}{R^{0{.}5}} \label{eq:n5}.
\end{equation}

This relation is obeyed more closely for both galaxies if we correct the relationship
between $\left<n_e\right>$ and $R$ for the dependence of $\left<n_e\right>$ on the 
galactocentric radius, $r$, as described in section 3.2 below. Making this correction
gives adjusted values of -0.52, and -0.47 for the exponents in M51 and NGC 4449, 
respectively.

\subsection{Electron density as a function of galactocentric distance}
\begin{figure}
\centering
   \includegraphics[width=3.1in]{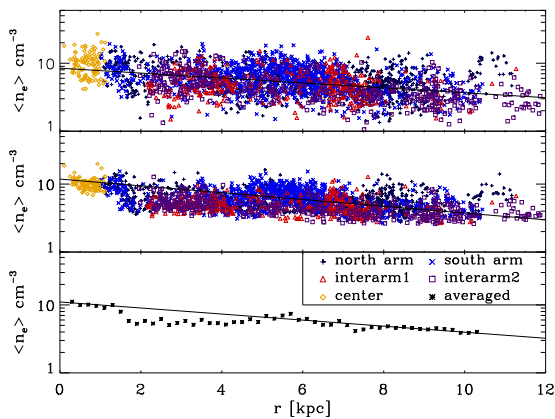} 
   \caption{Mean electron density ($\left<n_e\right>$) vs galactocentric 
    radius, $r$, for M51, distinguishing by the point styles those found in the arms, in the interarm zones, and within the central kpc. The points in the centre panel were found by applying a normalized correction factor to those in the upper panel, in order to take into account the variation of $\left<n_e\right>$ with region radius, $R$. In the lower panel we show the result of taking the median value of $\left<n_e\right>$ for the full sample of regiones within annuli of width 200 pc.
} \label{F:densidad_vs_distancia_m51}
\end{figure}

In the upper panel of Fig. \ref{F:densidad_vs_distancia_m51}  
we have plotted the mean electron 
density, $\left<n_e\right>$, against galactocentric radius, $r$, for M51, including all the regions, 
but distinguishing by point styles among those found in different large scale 
features of the galaxy: arms, interarm zones, and the central kpc. The 
outstanding feature of this plot is the single straight line fit (which is an 
exponential, as the electron density scale is logarithmic), which has the form:

\begin{equation}
\left<n_e\right> = {\left<n_e\right>_o}\; e^{-r/h}, \label{eq:n6}
\end{equation}

\noindent where $\left<n_e\right>_o$, with value 10$\pm$1 cm$^{-3}$ is a central value of electron density 
and $h$ is a scale length, which takes the value 10$\pm$1.0 kpc. 
We have limited the fit to galactocentric radii within $r$ = 10.4 kpc, because beyond 
this radius there is an abrupt local rise and fall in $\left<n_e\right>$, attributable to the 
effect of the interacting neighbour galaxy NGC 5195. 
We can make 
separate estimates for the constants $\left<n_e\right>_o$  and $h$ in the arms and in the 
interarm zones. For the arms this gives us $\left<n_e\right>_o$ = 10 cm$^{-3}$ and $h$ = 11 kpc,
while for the interarm zones we find $\left<n_e\right>_o$ = 8 cm$^{-3}$, and $h$ = 11 kpc . 


Since we know that in the upper panel of Fig. \ref{F:densidad_vs_distancia_m51} much of the apparent scatter is due to the (inverse) 
dependence of $\left<n_e\right>$ on the radii of the individual \hii{} regions, we next opted to 
apply a normalized correction factor for this effect to all the data, and replot the 
figure, yielding the central panel of Fig. \ref{F:densidad_vs_distancia_m51}, %
in which the 
(exponential) linear fit is clearly much improved. To make the fit we have chosen 
to leave out the zones from 1.4 to 4.6 kpc from the center, and from 10.4 to 11.5 kpc 
from the center, and now find values of  $\left<n_e\right>_o$ and $h$ of 12 cm$^{-3}$ and 9 kpc 
respectively. 
It is notable that the scale 
length found for the neutral atomic hydrogen component of M51 by \citet{tilanus91} 
is 9.1 kpc, on which we will comment further below. 
We should point out that the zone between 
the center of the galaxy and a radius of 1.4 kpc does fit the exponential for the 
disc as a whole. 

To take a look at the global behavior we have taken averages of all the regions within 
annuli of width 200 pc and have plotted, in the lower panel of Fig. \ref{F:densidad_vs_distancia_m51}
these values against 
galactocentric radius. This shows the overall trend in $\left<n_e\right>$, and we can pick out the 
general decline with radius, but note a plateau between 1.4 kpc and 4.6 kpc.
If we omit the 
points in this interval from those entering the exponential fit, we obtain values 
for $\left<n_e\right>_o$ and $h$, of 11 cm$^{-3}$ and 9 kpc respectively.

\begin{figure}
   \centering
     \includegraphics[width=3.1in]{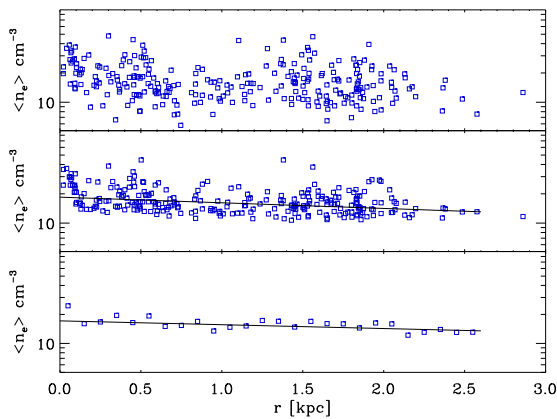} 
     \caption{Mean electron densities ($\left<n_e\right>$) for the measured \hi{} regions NGC 4449. In the upper panel $\left<n_e\right>$ 
   is plotted against galactocentric radius, in the centre 
panel $\left<n_e\right>$ has been modified by a correction factor to take into account the variation with the radii $R$ of the individual regions (a $1/R^{0.5}$ dependence), and in the lower panel we plot the median values of all the regions sampled within annuli of 100 pc width.}
   \label{F:densidad_vs_distancia_ngc4449}
\end{figure}   

To compute the effect of the temperature gradient on the determination of $\left<n_e\right>$ 
and of its gradient, we used data from \citet{bresolin04}
giving the O abundance gradient, and the widely used CLOUDY model \citep{ferland98}
to derive the effective temperature gradient, which we then incorporated into the expression

\begin{equation}
\left<n_e\right> = 2{.}3 \times 10^{-18} \sqrt{\frac{L({\ha{}})\,T^{0.91}} {R^3}} cm^{-3},\label{eq:n8}
\end{equation}

\noindent  and recalculated the electron densities. This introduced a change in the exponent of the relation 
between $\left<n_e\right>$ and radius from -0.56 to -0.55, and in the scale length from 9.2 to 9.9 
kpc. These are the values represented in Fig. \ref{F:densidad_vs_distancia_m51}. 

Here we should point out the similarity of the lower panel of Fig. \ref{F:densidad_vs_distancia_m51}
with the radial plot of the 
azimuthally averaged \hi{} column density in \citet{tilanus91}.  

General expressions for $\left<n_e\right>$ are then:

\begin{equation}
\left<n_e\right> = \left\{ 
\begin{array}{p{15mm}l}
\multicolumn{2}{l}{ 45{.}8\; e^{-r/10}\;R^{-0{.}55} cm^{-3};} \\
    \, &  r < 1{.}4\; \textrm{kpc}\; \textrm{\&} \; r > 4{.}6\; \textrm{kpc} \\ 
\multicolumn{2}{l}{ 28{.}9 \;R^{-0{.}55} cm^{-3};}  \\
    \, &  r \in [1{.}4,4{.}6]\; \textrm{kpc}.
\end{array}\right.  \label{eq:n7}
\end{equation}

The results for NGC 4449 are summarized in Fig. \ref{F:densidad_vs_distancia_ngc4449}. 
In the upper panel of Fig. \ref{F:densidad_vs_distancia_ngc4449} we have plotted $\left<n_e\right>$ against galactocentric radius for 
all our measured regions, in the central panel we present the data 
normalized to take into account the $1/R^{0.5}$ relation explained above, while in the lower panel 
we have averaged the points in the central panel in annuli of width 200 pc.
In the latter two panels the linear fits show the exponential radial decline in mean 
electron density, which is fitted by the whole galaxy outside the central 120 pc. Using the 
best fit in the lower panel of Fig. \ref{F:densidad_vs_distancia_ngc4449} and normalizing, $\left<n_e\right>$  is given by:

\begin{equation}
\left<n_e\right> = {45{.}7}\; e^{-r/11}\;R^{-0{.}45}. \label{eq:dens3}
\end{equation}

\noindent This scale length here is 11 kpc which is much bigger than the optical radius of the galaxy. However, 
\citet{hunter98} observed an \hi{}  disc with a central dense 
component some 4.5 kpc in radius embedded in a much larger elliptical component with a major axis 
limiting radius of 18 kpc. Taken our cue from M51, which has an \hi{} scale length of 9 kpc and a limiting \hi{} 
radius of 15 kpc \citep{meijerink05}, we can claim here that the \hii{} region scale length and the \hi{} scale 
length of NGC 4449 
are comparable, consistent with the scenario where the electron densities in the \hii{} regions follow the 
column density of the \hi{} gas, as in M51.

\subsection{The mean electron density  as a function of position in the spiral arms of M51}
\begin{figure}
\centering
   \includegraphics[width=3.1in]{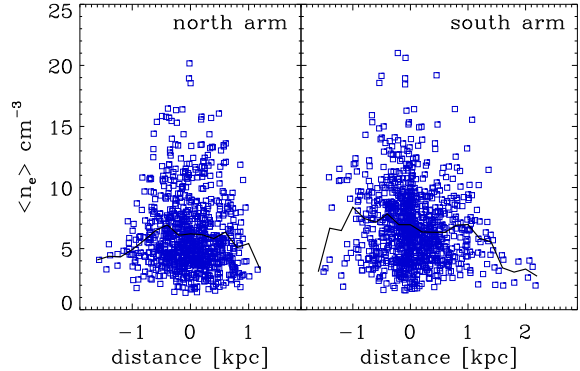} 
   \caption{ The mean electron densities of the HII 
regions in M51 plotted against distance 
from the central ridge line of the arms in M51 (squares). Also shown is the variation of the 
average value  in 200 pc bins of these densities as a function of this distance (thick lines) 
for both arms.} \label{F:densidad_vs_radiales}
\end{figure}

Finally in this section we show in Fig. \ref{F:densidad_vs_radiales} the electron density 
in the spiral arms of M51 as a function of the position relative to the central 
ridge line of the arms. The ridge lines of the 
arms, used to define this relative position, were measured using the continuum image 
through the F814W filter. The abscissa in these plots is the distance from the ridge line 
to the regions.
We can see that the mean electron density 
reaches peak values along the ridge line, which is not surprising, but taking the complete 
populations of \hii{} regions we can see that the systematic increase of density towards 
the ridge line is a property of the upper envelopes of the plots for the two arms. There 
are many regions close to the ridge line with 
relatively low densities, though no high density regions towards the edges of the arms.

\section{Conclusions}\label{sec:conclusions}
We have used the \ha{} luminosities and measured radii of the populations of 
\hii{} regions in M51 and NGC 4449, from HST images, to derive their mean electron 
densities, aware that their inhomogeneous structures imply that the mean 
electron density does not describe either the in situ densities of their 
strongly emitting clumps, or the densities of their more tenuous interclump 
gas, but is a luminosity weighted mean over the \hii{} region volume. We have 
found two functional dependences of this mean electron density parameter, valid 
for both galaxies: it varies inversely proportionally to the square root of the 
radius of an \hii{} region, and it falls exponentially with galactocentric radius. 
The former relation must depend on the effects of the energy outflow from the 
massive ionizing stars on their environment. As far as we are aware it does not 
correspond to the predictions of any previously published specific model 
(\citet{scoville01} show a graph for M51 which could be used to infer a 
similar result, but did not derive an explicit relation for these variables), and 
sets interesting constraints on the physics of OB cluster formation and the
 subsequent interaction with the cluster environment. The latter relation, with 
the scale length for the electron density comparable with that for the \hi{} column 
density in galaxies of very different sizes and types (M51 is a Sbc while 
NGC 4449 is an IBm), demonstrates that, globally, an \hii{} region is in 
quasi-pressure equilibrium with its surrounding gas. The quantitative differences 
between the density of arm and interarm regions are not great, but also reflect 
this quasi-equilibrium, since the values are systematically somewhat lower for 
the interarm regions. Finally the distribution of mean electron densities across 
a spiral arm has an upper envelope peaking along the center, the ridge line of 
the arm, but values well below the envelope are distributed fairly uniformly
 across the arm. These relations should be of value for improving physical 
descriptions of \hii{} regions as zones of interaction between hot stars and their 
surrounding gas.

\end{document}